\documentclass[reprint,amsmath,amssymb,aps,]{revtex4-1}
\usepackage{graphicx}
\usepackage{dcolumn}
\usepackage{amsmath}
\usepackage{amssymb}
\usepackage{bm}

\begin{document}
	
	\title{ Ohmic Reservoir-based non-Markovianity and Quantum Speed Limit Time}
	\author{Hong-Mei Zou\textsuperscript{1}}%
	\email{zhmzc1997@hunnu.edu.cn}
	\author{Rongfang Liu\textsuperscript{1}}
	\author{Dan Long\textsuperscript{1}}
	\author{Jianhe Yang\textsuperscript{1}}
	\author{Danping Lin\textsuperscript{2}}%
	\affiliation{\textsuperscript{1}Synergetic Innovation Center for Quantum Effects and Application, Key Laboratory of Low-dimensional Quantum Structures and Quantum Control of Ministry of Education, School of Physics and Electronics, Hunan Normal University, Changsha, 410081, People's Republic of China.\\
		\textsuperscript{2} Faculty of Science, Guilin University of Aerospace Technology, Guilin 541004,
		People's Republic of China.}%
	
	\begin{abstract}
		We study the non-Markovianity and quantum speedup of a two-level atom (quantum system of interest) in a dissipative Jaynes-Cumming model, where the atom is embedded in a single-mode cavity, which is leaky being coupled to an external reservoir with Ohmic spectral density. We obtain the non-Markovianity characterized by using the probability of the atomic excited state and the negative decoherence rate in the time-local master equation. We also calculate  the quantum speed limit time (QSLT) of the evolution process of the atom. The results show that, the atom-cavity coupling is the main physical reasons of the transition from Markovian to non-Markovian dynamics and the transition from no speedup to speedup process, and the critical value of this sudden transition only depends on the Ohmicity parameter. The atom-cavity coupling and the appropriate reservoir parameters can effectively improve the non-Markovianity in the dynamics process and speed up the evolution of the atom. Moreover, the initial non-Markovian dynamics first turns into Markovian and then back to non-Markovian with increasing the atom-cavity coupling under certain condition. Finally, the physical interpretation is provided.
		\begin{description}
			\item[PACS numbers]
			03.65.Yz, 03.67.Lx, 42.50.-p, 42.50.Pq.
			\item[Keywords]
			Quantum Speed Limit Time, non-Markovianity, dissipative cavity
		\end{description}
	\end{abstract}

	\maketitle
	
	\section{Introduction}
	As we known, the decoherence effect and the energy dissipation caused by coupling of system-environment will bring remarkable influences on the dynamical behaviour of the open system. The evolution process of the open system is Markovian if a quantum system is weakly coupled to a memoryless environment, while the evolution process is non-Markovian if a quantum system is strong coupled to a memory environment \cite{Davies,Lindblad,Kossakowski,Gorini,Verstraete,Rivas0}. The non-Markovian effect in the dynamics process can be described by the non-Markovianity. Many efforts have been made to define non-Markovianity, to measure it, and to take advantage of it \cite{Wolf1,Breuer1,Rivas1,S.Luo2,Zeng,He,Pineda,Poggi,He2014a,McCloskey2014a,Lorenzo2013a,Mortezapour2017a,Liu2018a,Gholipour2020a,RanganiJahromi2020,Jahromi2019a,Strasberg2019a,He2017a,Dhar2015a}. For examples, the non-markovianity was quantified by correlations in Ref. \cite{S.Luo2}, the authors in \cite{RanganiJahromi2020,He2017a} studied the measurement of non-Markovianity and Paternostro's team studied geometrical characterization of non-Markovianity \cite{Lorenzo2013a}. In recent years, the research on non-Markovianity in the dynamics process of the open system has attracted the attention of the community, both theoretically and experimentally \cite{Vega,Breuer0,Davalos1,ShangYu}.
	
	On the other hand, quantum speed limit (QSL) has been considered a purely quantum phenomenon with no corresponding concept in classical mechanics, which sets a bound on the maximal evolution velocity that a quantum system needs to evolve between two distinguishable states. Driving a given initial state to a target state at the maximal evolution speed is one of fundamental and important tasks of quantum physics, thus QSL plays a significant role in the fields of quantum computation, quantum metrology, and so on \cite{Anandan1,Vaidman1,S.Luo1,Lloyd1,Giovannetti}. The minimal evolution time between two distinguishable states of a quantum system is defined as the quantum speed limit time(QSLT) \cite{Jones1,Zwierz1,Z.Y.Xu1,Xiangji1,Francesco1}. For closed systems, the QSLT is defined by $\tau=\max\{\frac{\pi\hbar}{2\Delta E},\frac{\pi\hbar}{2E}\}$, where $\Delta E$ in Mandelstam-Tamm(MT) bound \cite{Mandelstam} and $E$ in  Margolus-Levitin(ML) bound \cite{Margolus} are the fluctuation and the mean value of the initial-state energy, respectively. For open systems, Deffner and Lutz obtained the unified bound of QSLT from the MT and ML types by using the Bures angle and showed that the non-Markovian effects could speed up the quantum evolution  \cite{Deffner1}.  In recent years, many efforts have been made in the study of QSLT of an open system \cite{Taddei1,delCampo1,Shao-xiong,Zhang1,S.-X.Wu1,Liu,Cianciaruso1,Ahansaz2019}.
	
	In addition, much valuable effort have also been devoted to the relationships between the non-Markovianity and the QSL \cite{Xu2014a,HaiBin,Mirkin2016a,Wang1,Ahansaz2019,Deffner2013a,Xu2018a}, such as quantum speedup in a memory environment \cite{Xu2014a}, quantum speedup in open quantum systems \cite{HaiBin,Mirkin2016a} and the relationship between the quantum speedup and the formation of a system-environment bound state \cite{Wang1,Ahansaz2019}. The authors in \cite{YingJie} found that a classical field can effectively regulate the non-Markovianity and the QSLT of an open qubit. Namely, the strong coupling of qubit-environment and an external classical field can all realize the transformation from Markovian to non-Markovian dynamics and the speedup evolution of the system. In these studies mentioned above, the environment is usually at zero temperature and has generally the Lorentzian spectral density. 
	
	In Ref. \cite{Zou5}, we have studied the QSLT and the non-Markovianity of the atom in Jaynes-Cummings model coupling with the Lorentzian reservoir and the Ohmic reservoir with a Lorentz–Drude cutoff function, respectively, and the reservoir is at zero temperature, and we characterized the non-Markovianity by using the positive derivative of the trace distance. However, in this paper, we focus on the Ohmic reservoir and the master equation of the atom-cavity subsystem interacting with the reservoir at $T$ temperature, and we characterize the non-Markovianity by using the probability of the atomic excited state and the negative decoherence rate in the time-local master equation. The results show that the atom-cavity coupling and the appropriate reservoir parameters can improve the non-Markovianity in the dynamics process and accelerate the evolution of the atom. 
	
	The outline of the paper is the following. In Section II, we describe a physical model. In Section III, we introduce the non-Markovianity and the quantum speed limit time. Results and discussions are provided in Section IV. Finally, we give a brief summary in Section V.
	
	\section{ Physical model}
	We consider a dissipative Jaynes-Cummings model \cite{Jaynes,Shore}, namely, an atom is in a leaky cavity that the leakage is usually modelled by coupling of the cavity mode to the bosonic modes of the reservoir. The Hamiltonian of the total system is given by ($\hbar =1$)
	\begin{equation} \label{EB201}
	\hat{H}=\hat{H}_{JC}+\hat{H}_{CR}
	\end{equation}
	here 
	\begin{equation}\label{EB202}
	\hat{H}_{JC}=\frac{1}{2}\omega _{0}\hat{\sigma}_{z}+\omega _{0}\hat{a}^{\dag }\hat{a}+\Omega (\hat{a}\hat{\sigma}_{+}+\hat{a}^{\dag }\hat{\sigma}_{-})    
	\end{equation}%
	and
	\begin{equation}\label{EB2022}
	\hat{H}_{CR}=\sum_{k}\omega_{k}\hat{b}_{k}^{\dag}\hat{b}_{k}+(\hat{a}+\hat{a}^{\dag})\sum_{k}g_{k}(\hat{b}_{k}^{\dag}+\hat{b}_{k})
	\end{equation}%
	 where the atomic transition frequency is $\omega _{0}$ and the Pauli matrices of the atom are  $\hat{\sigma}_{z}$ and $\hat{\sigma}_{\pm }$. $\hat{a}^{\dag }$($\hat{a}$) and $\hat{b}_{k}^{\dag }$ ($\hat{b}_{k}$) express the creation(annihilation) operators of the cavity and the $k$-th mode of reservoir with the frequency $\omega _{k}$, respectively. $\Omega $ and $g_{k}$ are the coupling strength of the atom-cavity and the cavity-reservoir, respectively.
	
	In this work, we suppose that the total number of excitations is $n=1$ in the total system. The eigenstates and eigenvalues of the Hamiltonian $\hat{H}_{JC}$ are given by $|\varphi _{1,\pm }\rangle =\frac{1}{\sqrt{2}}(|1,g\rangle \pm |0,e\rangle )$ and $E_{1,\pm } =\frac{1}{2}\omega_{0}\pm\Omega$ for $n=1$, while the ground state and the corresponding energy eigenvalue are $|\varphi _{0}\rangle =|0,g\rangle $ and $E_{0} =-\frac{1}{2}\omega_{0}$. Then we assume that $\hat{A}_{1}^{+}=|\varphi _{1,-}\rangle \langle \varphi _{0}|$ and $\hat{A}_{1}^{-}=|\varphi _{0}\rangle \langle \varphi _{1,-}|$ are the jump operators between $|\varphi _{1,-}\rangle $ and $|\varphi _{0}\rangle $, and $\hat{A}_{2}^{+}=|\varphi _{1,+}\rangle \langle \varphi _{0}|$ and $\hat{A}_{2}^{-}=|\varphi _{0}\rangle \langle \varphi _{1,+}|$ are the jump operators between $|\varphi _{1,+}\rangle $ and $|\varphi _{0}\rangle $. Performing the Born-Markov and the rotating wave approximations, tracing out the freedom degrees of the reservoir in the interaction picture and then going back to the Schr\"{o}dinger picture, we can obtain the master equation for the atom-cavity subsystem interacting with the reservoir at $T$ temperature as follows \cite{Scala1,Zou3}:
	\begin{equation}\label{EB2031}
	\begin{split}
	\frac{d}{dt}\varrho (t) &=-i[\hat{H}_{JC},\varrho(t)]\\
	&+\frac{1}{2}\gamma(\omega_{1},t)(\hat{A}_{1}^{-}\varrho (t)\hat{A}_{1}^{+}-\frac{1}{2}\{\hat{A}_{1}^{+}\hat{A}_{1}^{-},\varrho (t)\}) \\
	&+\frac{1}{2}\gamma(\omega_{2},t)(\hat{A}_{2}^{-}\varrho (t)\hat{A}_{2}^{+}-\frac{	1}{2}\{\hat{A}_{2}^{+}\hat{A}_{2}^{-},\varrho (t)\}) \\
	&+\frac{1}{2}\gamma(-\omega_{1},t)(\hat{A}_{1}^{+}\varrho (t)\hat{A}_{1}^{-}-\frac{1}{2}\{\hat{A}_{1}^{-}\hat{A}_{1}^{+},\varrho (t)\}) \\
	&+\frac{1}{2}\gamma(-\omega_{2},t)(\hat{A}_{2}^{+}\varrho (t)\hat{A}_{2}^{-}-\frac{	1}{2}\{\hat{A}_{2}^{-}\hat{A}_{2}^{+},\varrho (t)\})
	\end{split}
	\end{equation}%
	here $\omega_{1,2}=\omega _{0}\mp\Omega $ is the transition frequency of the dressed-states $|\varphi _{1,\mp}\rangle \leftrightarrow |\varphi _{0}\rangle $. $\gamma(\omega_{1},t)$ and $\gamma(\omega_{2},t)$ are the time dependent decay rates for $|\varphi _{1,-}\rangle $ and $|\varphi _{1,+}\rangle $, respectively, i.e.
	\begin{equation}\label{EB207}
	\gamma({\omega_{j},t})=2\Re[{\int_{0}^{t}d\tau \int_{-\infty }^{+\infty
		}d\omega e^{i(\omega _{j}-\omega)\tau }J(\omega)}]
	\end{equation}%
	in which $J(\omega)$ is the spectral density of the reservoir and
	\begin{equation}\label{EB2032}
	\begin{split}
	\gamma(-\omega_{j},t) &=\exp(-\frac{\omega_{j}}{k_{B}T})\gamma(\omega_{j},t)
	\end{split}
	\end{equation}%
	For simplicity, we only discuss the reservoir at zero temperature \cite{Spohn1978} in the following. Eq.~(\ref{EB2031}) becomes 
	\begin{equation}\label{EB203}
	\begin{split}
	\frac{d}{dt}\varrho (t) &=-i[\hat{H}_{JC},\varrho(t)]\\
	&+\frac{1}{2}\gamma(\omega_{1},t)(\hat{A}_{1}^{-}\varrho (t)\hat{A}_{1}^{+}-\frac{1}{2}\{\hat{A}_{1}^{+}\hat{A}_{1}^{-},\varrho (t)\}) \\
	&+\frac{1}{2}\gamma(\omega_{2},t)(\hat{A}_{2}^{-}\varrho (t)\hat{A}_{2}^{+}-\frac{	1}{2}\{\hat{A}_{2}^{+}\hat{A}_{2}^{-},\varrho (t)\}) 
	\end{split}
	\end{equation}%
			
	We can acquire an analytical solution of the density operator $\varrho (t)$ in the dressed-state basis $\{|\varphi _{0}\rangle,|\varphi _{1,-}\rangle,|\varphi _{1,+}\rangle \}$ from Eq.~(\ref{EB203}), then the density matrix $\rho (t)$ of the atom in the standard basis $\{|e\rangle ,|g\rangle \}$ is also obtained by means of the representation transformation and taking a partial trace over the freedom degrees of the cavity. Suppose the initial state is $\{\rho _{11}(0),\rho_{10}(0),\rho _{01}(0),\rho _{00}(0)\}$, the density matrix $\rho (t)$ \cite{Zou5} of the atom at all time $t$ is expressed as 
	\begin{equation}\label{EB204}
	\rho (t)=\left( 
	\begin{array}{cc}
	|p(t)|^{2}\rho _{11}(0) & p(t)\rho _{10}(0)  \\ 
	p(t)^{\ast }\rho _{01}(0) & 1-|p(t)|^{2}\rho _{11}(0) 
	\end{array}%
	\right) 
	\end{equation}%
	where the probability amplitude $p(t)$ can be given by 
	\begin{equation}\label{EB205}
	p(t)=\frac{1}{2}\sum_{j=1}^{2}e^{-i\omega _{j}t}e^{-\frac{1}{4}\beta _{j}}
	\end{equation}%
	here
	 \begin{equation}\label{EB206}
	 \beta _{j}=\int_{0}^{t}\gamma(\omega _{j},t^{\prime })dt^{\prime }
	 \end{equation}%
	Considering the structured reservoir with an Ohmic spectral density 
	\begin{equation}  \label{EB208}
	J(\omega)=\eta\omega^{s}\omega_{c}^{1-s} e^{-\omega/\omega_{c}}
	\end{equation}
	where $s$ is the Ohmicity parameter, which moves the spectrum from sub-Ohmic ($0<s<1$) to Ohmic if ($s=1$) and super-Ohmic ($s>1$) regimes \cite{Ming-Liang,Leggett,Benedetti}. $\omega_{c}$ and $\eta$ being the cut-off frequency and the dimensionless coupling constant, which are related to the reservoir correlation time $\tau_{B}$ and the relaxation time $\tau_{R}$ (over which the state of the system changes in the Markovian limit of a flat spectrum) by $\tau_{B}\approx\omega_{c}^{-1}$ and $\tau_{R}\approx\eta^{-1}$. $\omega_{c}<\omega_{0}$ implies that the spectrum of the reservoir does not completely overlap with the frequency of the cavity, that is, the reservoir is effectively adiabatic, so that the evolution behaviour of the system is essentially non-Markovian. While $\omega_{c}>\omega_{0}$ indicates the converse case, which the quantum information is quickly dissipated, the evolution behaviour of the system is Markovian. The smaller the value of $\eta$ is, the longer the reservoir correlation time is, and the more obvious the non-Markovian effect is \cite{Zou4,Eckel,CuiW}. 
	
	Common values of $s$ are $\frac{1}{2}$, 1 and 3, inserting Eq.~(\ref{EB208}) into Eq.~(\ref{EB207}), $\gamma({\omega_{j},t})$ is written as 
	\begin{equation}\label{EB209}
	s=\frac{1}{2}:\gamma({\omega_{j},t})=-\frac{2\eta \omega _{c}\sqrt{\pi}}{(1+\omega_{c}^{2}t^{2})^{\frac{1}{4}}}\sin(\omega _{j}t-\frac{\alpha _{0}}{2})
	\end{equation}%
	\begin{equation}\label{EB210}
	s=1:\gamma({\omega_{j},t})=-\frac{2\eta \omega _{c}}{(1+\omega_{c}^{2}t^{2})^{\frac{1}{2}}}\sin(\omega _{j}t-\alpha _{0})
	\end{equation}%
	\begin{equation}\label{EB211}
	\begin{split}
	s=3:\gamma({\omega_{j},t}) &=-\frac{2\eta \omega _{j}^{2}}{\omega_{c}(1+\omega_{c}^{2}t^{2})^{\frac{1}{2}}}\sin(\omega _{j}t-\alpha _{0})  \\
	&-\frac{2\eta \omega _{j}}{(1+\omega_{c}^{2}t^{2})}\sin(\omega_{j}t-2\alpha _{0}) \\
	&-\frac{4\eta \omega _{c}}{\omega _{c}(1+\omega _{c}^{2}t^{2})^{\frac{3}{2}}}\sin(\omega _{j}t-3\alpha _{0})
	\end{split}
	\end{equation}%
	with $\alpha _{0}=\arctan(\omega _{c}t)$.
	
	We can not get the analytical expressions for $\beta_{j}$ from Eq.~(\ref{EB206}) and Eqs.~(\ref{EB209})-(\ref{EB211}), but we can calculate mathematically $\beta_{j}$ for the sub-Ohmic, Ohmic and super-Ohmic spectra, respectively.
	
	In view of Eq.~(\ref{EB204}), we can also write a time-local master equation \cite{Breuer} for the density operator $\rho (t)$ as 
	\begin{equation}\label{EB212}
	\begin{split}
	\frac{d}{dt}\rho (t) &=\mathcal{L}\rho (t) \\
	&=-\frac{i}{2}S(t)[\hat{\sigma}_{+}\hat{\sigma}_{-},\rho (t)]+\varGamma (t)\{\hat{\sigma}_{-}\rho(t)\hat{\sigma}_{+} \\
	&-\frac{1}{2}\hat{\sigma}_{+}\hat{\sigma}_{-}\rho (t)-\frac{1}{2}\rho (t)\hat{\sigma}_{+}\hat{\sigma}_{-}\}
	\end{split}
	\end{equation}%
	where the Lamb frequency shift $S(t)$ and the decoherence rate $\varGamma(t)$ can be respectively expressed as
	\begin{equation}\label{EB213}
	S(t)=-2\Im \lbrack \frac{\dot{p}(t)}{p(t)}]   
	\end{equation}%
	and
	\begin{equation}\label{EB214}
	\varGamma (t)=-2\Re\lbrack \frac{\dot{p}(t)}{p(t)}]
	\end{equation}%
	$S(t)$ describes the contribution from the unitary part of the evolution under dynamical decoherence.  $\varGamma(t)$ characterizes the dissipation and the feedback of the information of the system. $\varGamma(t)>0$ indicates that quantum information flows from the system to its environment, i.e. Markovian process. $\varGamma(t)<0$ expresses that quantum information flows back from its environment to the system, i.e. non-Markovian process.
	
	\section{ Non-Markovianity and Quantum speed limit time}
	\subsection{ Non-Markovianity}
	In the dynamics process of an open system, the non-Markovianity can describe the total backflow of information to the system from its environment. Among the different measurement of the non-Markovianity, the method based on the time rate of change of the trace distance is more commonly used at present. The trace distance between $\rho _{1}(t)$ and $\rho _{2}(t)$ is defined as $\mathcal{D}(\rho_{1}(t),\rho_{2}(t))=\frac{1}{2}Tr\Vert \rho _{1}(t)-\rho_{2}(t)\Vert $, which expresses the distinguishability between the two states $\rho _{1,2}(t)$ evolving from their respective initial forms $\rho_{1,2}(0) $ \cite{Breuer1}. The time rate of change of the trace distance can be expressed as $\sigma (t,\rho _{1,2}(0))=\frac{d}{dt}\mathcal{D}(\rho_{1}(t),\rho _{2}(t))$. $\sigma (t,\rho _{1,2}(0))<0$ indicates that $\mathcal{D}(\rho _{1}(t),\rho _{2}(t))$ decreases with time because the information flows irreversibly from the system to the environment, $\sigma(t,\rho _{1,2}(0))>0$ shows that $\mathcal{D}(\rho _{1}(t),\rho _{2}(t))$ is no longer decreasing monotonously because the information backflow from the environment to the system. The non-Markovianity can be calculated by $\mathcal{N}=\max_{\rho _{1,2}}\int_{\sigma >0}\sigma (t,\rho _{1,2}(0))dt$ \cite{HaiBin,Baumgratz}. If $\sigma (t,\rho_{1,2}(0))<0$, $\mathcal{N}=0$ and the dynamics process of the system is Markovian. If $\sigma (t,\rho _{1,2}(0))>0$, $\mathcal{N}>0$ and the dynamics process of the system is non-Markovian. 
	
	For the atom in Eq.~(\ref{EB204}), it has been proven that the optimal pair of initial states to maximize $\mathcal{N}$ are $\rho _{1}(0)=|e\rangle \langle e|$ and $\rho _{2}(0)=|g\rangle \langle g|$ \cite{Z.Y.Xu1,Deffner1}. Therefor the trace distance and its time rate of change are 
	\begin{equation}\label{EB301}
	\mathcal{D}(\rho _{1}(t),\rho_{2}(t))=|p(t)|^{2} 
	\end{equation}
	and
	\begin{equation}\label{EB3001}
	\sigma(t,\rho _{1,2}(0))=\frac{d}{dt}|p(t)|^{2} 
	\end{equation}
	From Eq.~(\ref{EB214}) and Eq.~(\ref{EB3001}), the decoherence rate $\varGamma(t)$ can be obtained
	\begin{equation}\label{EB3002}
	\varGamma(t)=-\frac{\sigma}{|p(t)|^{2}}
	\end{equation}
	
	Therefore, the non-Markovianity can be characterized by using the probability of the atomic excited state and the negative decoherence rate in the time-local master equation as 
	\begin{equation}\label{EB3003}
	\mathcal{N}=-\int_{\varGamma(t)<0}|p(t)|^{2}\varGamma(t)dt
	\end{equation}
	that is, there is non-Markovian in the dynamical process if the decoherence rate $\varGamma(t)$ is negative because the probability $|p(t)|^{2}$ is non negative. In the dissipative JC model, the quantum information will be exchanged between the reservoir with the cavity and between the cavity with the atom. Because we only care about the dynamics of the atom, both the cavity and its outside reservoir are regarded as the atomic environment. From Eqs.~(\ref{EB205}) and ~(\ref{EB3002})-(\ref{EB3003}), we can see that the non-Markovianity $\mathcal{N}$ is determined by all environment parameters (including the atom-cavity coupling $\Omega$, the cavity-reservoir coupling $\eta$, the cut-off frequency $\omega_{}c$ and the value of $s$). The non-Markovianity $\mathcal{N}$ is larger, the information from the environment feeding back to the atom is more. 
	
	\subsection{ Quantum speed limit time}
	The bound of the minimal evolution time from an initial state $\rho (0)$ to a final state $\rho (\tau )$ is  defined as the quantum speed limit time (QSLT) of a syetem, where $\tau $ is an actual evolution time. If the initial state is $\rho (0)=|\psi_{0}\rangle \langle \psi _{0}|$ and its target state $\rho (\tau )$ satisfies the master equation $\dot{\rho}(t)=\mathcal{L}\rho (t)$(see Eq.~(\ref{EB212})) with $\mathcal{L}$ being the positive generator of the dynamical semigroup, the QSLT can expressed as $\tau _{QSL}=\sin ^{2}\beta \lbrack \rho (0),\rho (\tau)]/\Lambda _{\tau }^{\infty }$ according to the unified lower bound derived by Deffner and Lutz, where $\beta \lbrack \rho (0),\rho (\tau )]=\arccos\sqrt{\langle \psi_{0}|\rho _{\tau }|\psi _{0}\rangle }$ indicates the Bures angle between $\rho (0)$ and $\rho (\tau )$, and $\Lambda _{\tau}^{\infty }=\tau ^{-1}\int_{0}^{\tau }\Vert \mathcal{L}\rho (t)\Vert dt$ with the operator norm $\Vert B\Vert $ equal to the largest eigenvalue of $\sqrt{B^{\dag }B}$ \cite{Deffner1}. When $\rho (0)=|e\rangle \langle e|$, we can obtain the QSLT from Eq.~(\ref{EB204}) as 
	\begin{equation}\label{EB303}
	\frac{\tau _{QSL}}{\tau }=\frac{1-|p(t)|^{2}}{\int_{0}^{\tau }\partial
		_{t}|p(t)|^{2}dt} 
	\end{equation}%
	
	For the dynamics process from $\rho (0)$ to $\rho (\tau )$, the non-Markovianity is also written as 
	\begin{equation}\label{EB302}
	\mathcal{N}=\frac{1}{2}[\int_{0}^{\tau }|\partial _{t}|p(t)|^{2}|dt+|p(\tau)|^{2}-1]
	\end{equation}%
	
	From Eqs.~(\ref{EB303})-~(\ref{EB302}), the relationship \cite{Z.Y.Xu1} between the QSLT and the non-Markovianity can be obtained as 
	\begin{equation}  \label{EB304}
	\frac{\tau_{QSL}}{\tau}=\frac{1-|p(\tau)|^{2}}{1-|p(\tau)|^{2}+2\mathcal{N}}
	\end{equation}
	Eq.~(\ref{EB304}) shows that the QSLT is equal to the actual evolution time when $\mathcal{N}=0$, but the QSLT is smaller than the actual evolution time when $\mathcal{N}>0$. That is, the non-Markovianity in the dynamics process can lead to the faster quantum evolution and the smaller QSLT.
	
	\section{Results and Discussions}
	In this section, we analyse the relations between the trace distance with its derivative, between the non-Markovianity with the decoherence rate and the derivative of the trace distance, as well as between the non-Markovianity with the quantum speed limit time. We also study the influence of the atom-cavity coupling and the reservoir parameters on the the non-Markovianity and the quantum speed limit time.
	
	In Fig.1, we draw the curve of the trace distance and its derivative, the decoherence rate and the non-Markovianity when $s=1$ (Ohmic spectrum), $\Omega=3\omega_{0}$ and $\frac{\omega_{c}}{\omega_{0}}=2$. We find that, the trace distance degenerates to zero from 1.0, and the derivative of the trace distance becomes negative and the decoherence rate simultaneously increases from zero. Then the trace distance again increases from zero, and the derivative of the trace distance becomes positive and the decoherence rate suddenly becomes negative at the same time. In addition, we can see that the non-Markovianity is equal to zero when the decoherence rate is positive (i.e. the derivative of the trace distance is negative), in which the quantum information flows from the system to its environment due to the dissipation of environment. The non-Markovianity is larger than zero when the decoherence rate is negative (i.e. the derivative of the trace distance is positive), where the quantum information flows back from its environment to the system because of the memory and feedback of environment. Therefore, once $\mathcal{D}(t)$ increases, the positive value of $\sigma(t)$ and the negative value of the decoherence rate will appear at the same time, the non-Markovianity in the dynamics process can be witnessed. 
	
	\begin{figure}[tbp]
		\includegraphics[width=7cm,height=5cm]{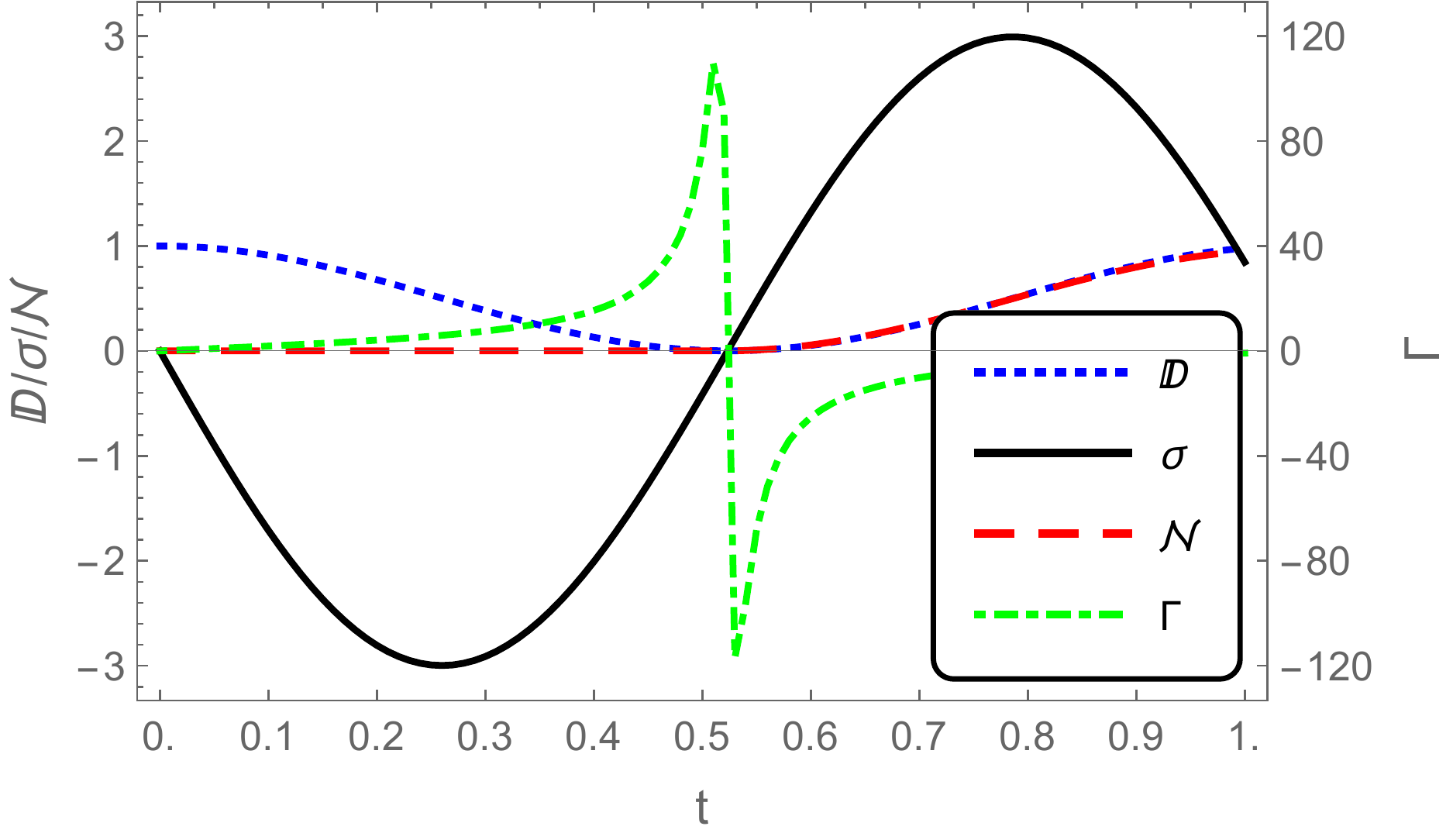} 
		\caption{(Color online) The dynamical curve of the trace distance and its derivative, the decoherence rate and the non-Markovianity when $s=1$ (Ohmic spectrum), $\Omega=3\omega_{0}$ and $\frac{\omega_{c}}{\omega_{0}}=2$. The left coordinate is $\mathcal{D}(t)$, $\sigma(t)$ and $\mathcal{N}(t)$ while the right coordinate is $\varGamma(t)$. The blue dotted line is the trace distance $\mathcal{D}(t)$, the black solid line is the derivative $\sigma(t)$ of the trace distance, the red dashed line is the non-Markovianity $\mathcal{N}$, and the green dotted-dashed line is the decoherence rate $\varGamma(t)$. The other parameters are $\eta=0.1$, $\omega_{0}=1$ and $\omega_{c}=2$.}
		\label{fig:1}
	\end{figure}
	
	Fig.2(a) shows the dynamical properties of the derivative of the trace distance under different atom-cavity coupling $\Omega$. $\sigma(t)$ changes from zero to negative when $\Omega=\omega_{0}$, thus the shaded area with positive $\sigma(t)$ is missing which means $\mathcal{D}(t)$ is nonincreasing during the whole evolution, shown as the green dotted line in Fig.2(a). $\sigma(t)$ changes from zero to negative and then again to zero when $\Omega=1.55\omega_{0}$, shown as the brown dashed line in Fig.2(a), the shaded area with positive $\sigma(t)$ is still zero but this is a threshold, that is, the shaded area with positive $\sigma(t)$ is appearing if $\Omega>1.55\omega_{0}$. When $\Omega=3\omega_{0}$, the red solid line changes from zero to negative and then to positive which means $\mathcal{D}(t)$ decreases first and then increases during the whole evolution, the shaded area with positive $\sigma(t)$ is $\int_{\sigma >0}\sigma (t,\rho _{1,2}(0))dt=0.945$, as shaded in Fig.2(a). The non-Markovianity $\mathcal{N}$ is plotted in Fig.2(b) versus $\Omega/\omega_{0}$. For the region with $\Omega<1.55\omega_{0}$, $\mathcal{N}$ is always zero, which means the derivative $\sigma(t)$ can never give a positive value, an example with $\Omega=\omega_{0}$ is the green dot ($\mathcal{N}=0$) in Fig.2(b) corresponding to the shaded area with positive $\sigma(t)$ of the green line in Fig.2(a). For the region with $\Omega>1.55\omega_{0}$, there always exists a positive value of $\mathcal{N}$, the red dot ($\mathcal{N}=0.948$) in Fig.2(b) epresents an example of $\Omega=3\omega_{0}$ which corresponds to the shaded area with the red line in Fig.2(a). The critical point with $\Omega=1.55\omega_{0}$ shows the situation of the transition from Markovianity to non-Markovianity, which the brown dot ($\mathcal{N}=0$) in Fig.2(b) corresponds to the shaded area with positive $\sigma(t)$ of the brown line in Fig.2(a). Namely, the atom-cavity coupling is the main physical reasons of the transition from Markovian to non-Markovian dynamics and enhancing the non-Markovianity in the dynamics process.
	
	\begin{figure}[tbp]
		\includegraphics[width=7cm,height=4.5cm]{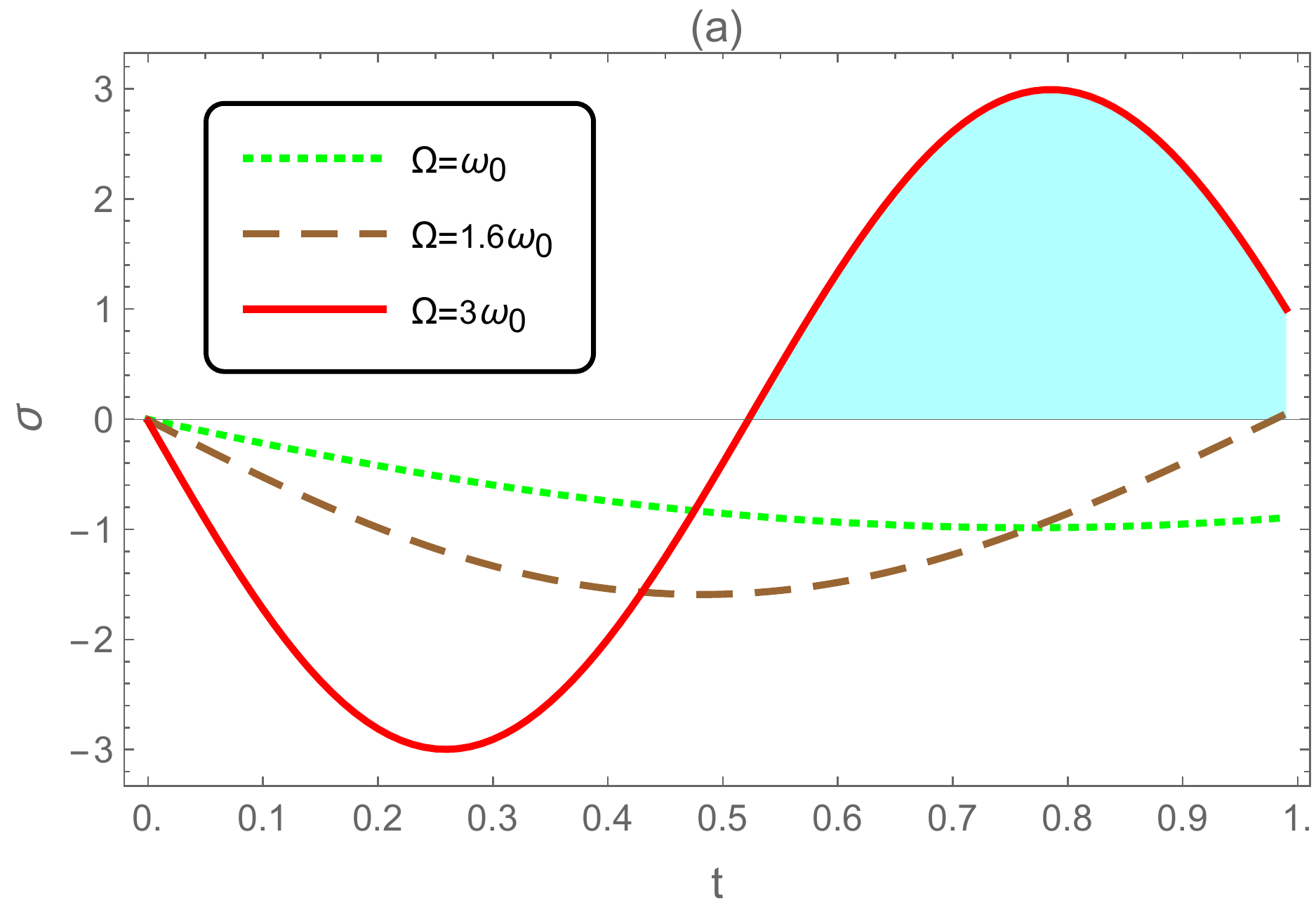} 
		\includegraphics[width=7cm,height=4.5cm]{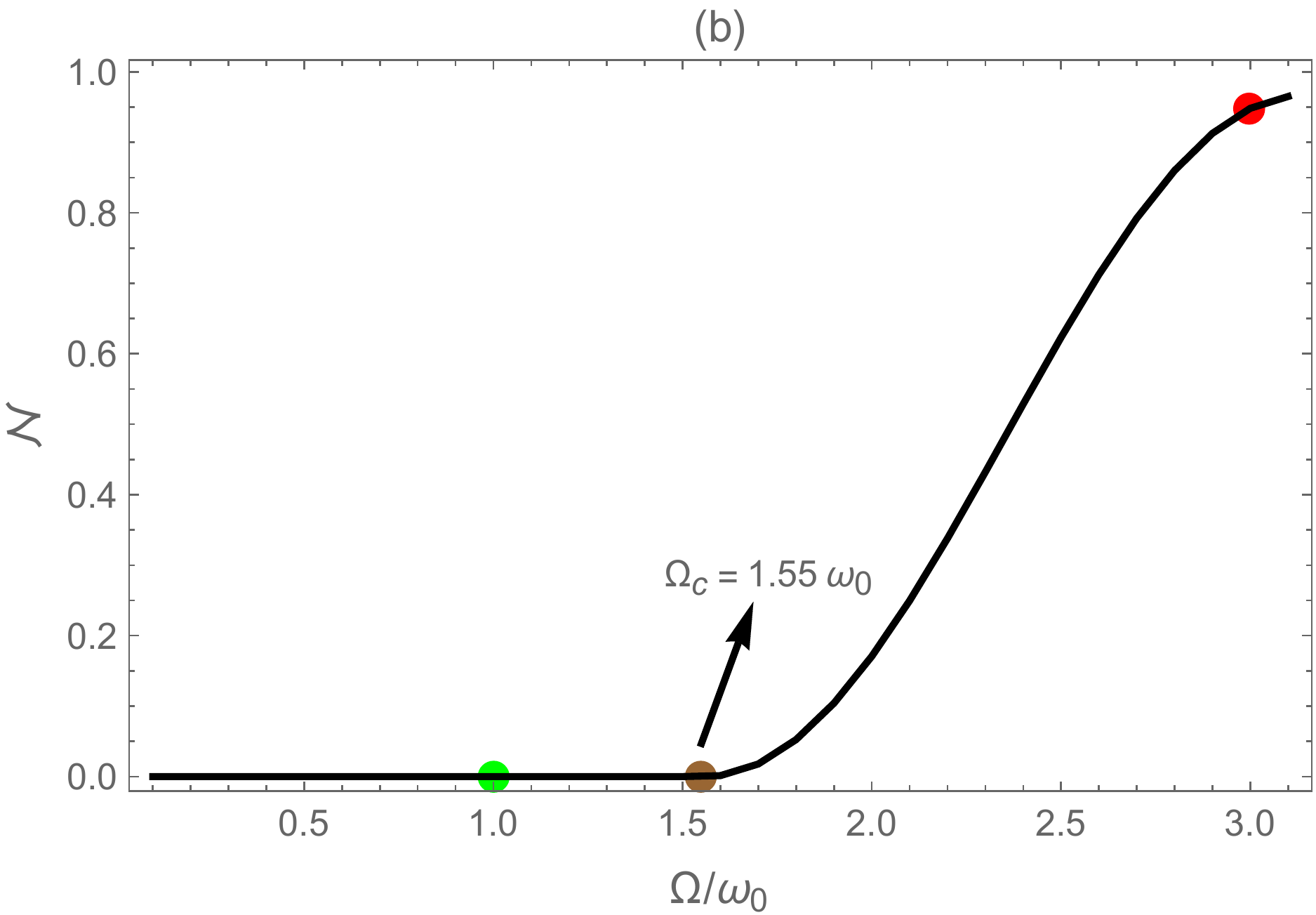} 
		\caption{(Color online)The dependence of the non-Markovianity $\mathcal{N}$ on the derivative $\sigma(t)$ of the trace distance and the coupling strength $\Omega$ when $s=1$ (Ohmic spectrum) and $\frac{\omega_{c}}{\omega_{0}}=2$. (a)The curves of the derivative $\sigma$ for different $\Omega$ values: the green dotted line is an example with $\Omega=\omega_{0}$, the brown dashed line indicates an example with $\Omega=1.55\omega_{0}$ and the red solid line represents $\Omega=3\omega_{0}$ where the positive region of $\sigma(t)$ is shaded. (b)Non-Markovianity as a function of the coupling strength $\Omega$, in which the green dot is corresponding to the green line in (a), the brown dot corresponds to the brown line in (a) which is the transition point from Markovian to non-Markovian dynamics,  and the red dot corresponds to the red line in (a). The other parameters are $\eta=0.1$, $\omega_{0}=1$ and $\omega_{c}=2$.}
		\label{fig:2}
	\end{figure}
	
	In Fig.3, we exhibit the curves of the non-Markovianity and the QSLT as functions of the coupling strength $\Omega$ when $s=1$ and $\frac{\omega_{c}}{\omega_{0}}=2$ for different coupling constant $\eta$, respectively. Fig.3(a) shows that $\mathcal{N}$ is always zero when $\Omega<\Omega_{c}$ and $\mathcal{N}$ will increase with $\Omega$ enlarging when $\Omega>\Omega_{c}$. Namely, there is a critical value $\Omega _{c}$ that $\mathcal{N}$ steeply increases from zero and the critical value is same for different coupling constant $\eta$. However, the increasing rate of $\mathcal{N}$ depends on the value of $\eta$, i.e., the smaller the coupling $\eta$, the stronger the non-Markovianity. The dependence of QSLT on the coupling $\Omega$ and $\eta$ is dotted in Fig.3(b), we find that $\tau_{QSLT}$ is always equal $\tau$ when $\Omega<1.55\omega_{0}$ and $\tau_{QSLT}$ will decrease with $\Omega$ enlarging when $\Omega>1.55\omega_{0}$. Namely, there is a critical value $\Omega _{c}$ of a sudden transition from no speedup to speedup and the critical value is same for different coupling constant $\eta$. But the decreasing rate of $\tau_{QSLT}$ depends on the value of $\eta$, i.e., the smaller coupling $\eta$ is corresponding to the more obvious speedup process. This shows that, in addition to the atom-cavity coupling $\Omega$, the cavity-reservoir coupling $\eta$ can also regulate the non-Markovianity in the dynamics process and the speedup evolution process of the atom.
	
	\begin{figure}[tbp]
		\includegraphics[width=7cm,height=4.5cm]{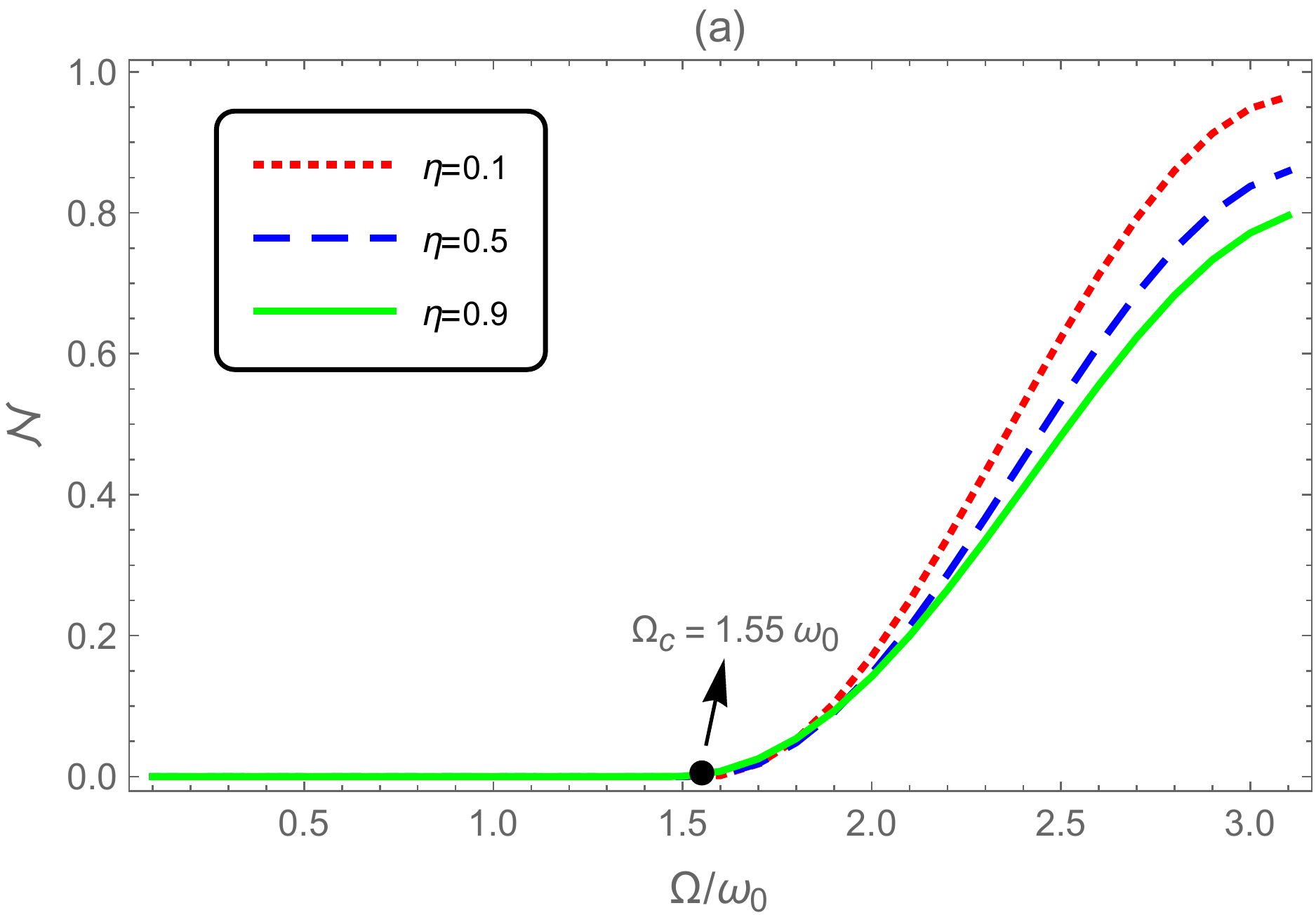} 
		\includegraphics[width=7cm,height=4.5cm]{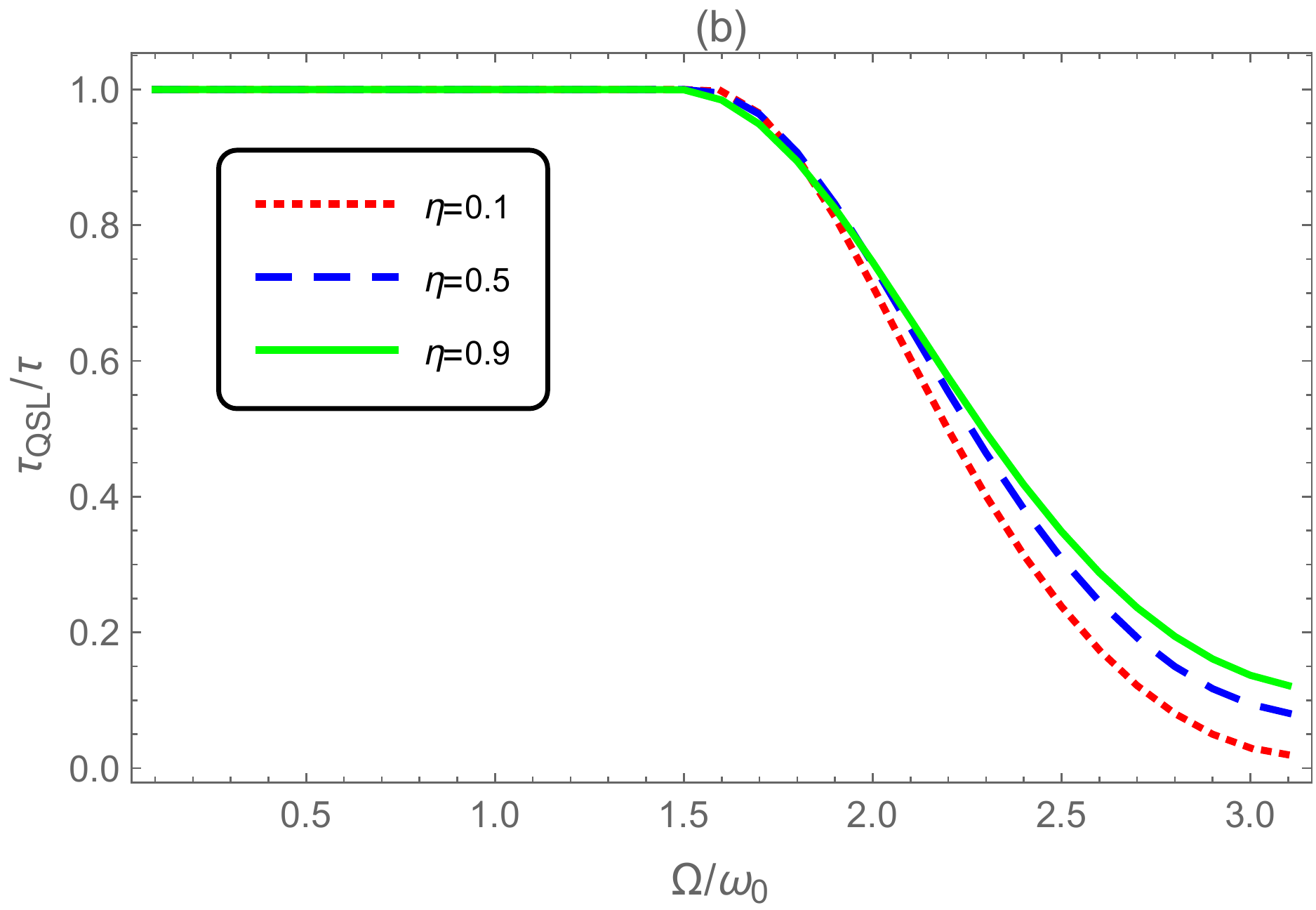}
		\caption{(Color online)Non-Markovianity and QSLT as a function of the coupling strength $\Omega$ when $s=1$ and $\frac{\omega_{c}}{\omega_{0}}=2$ for different coupling constant $\eta$, respectively. $\eta=0.1$, red dotted line; $\eta=0.5$, blue dashing line; $\eta=0.9$, green solid line. (a)Non-Markovianity as a function of $\Omega$; (b)QSLT as a function of $\Omega$. The other parameters are $\omega_{0}=1$ and $\omega_{c}=2$.}
		\label{fig:3}
	\end{figure}
	
	In Fig.4, we describe the dependence relation of the non-Markovianity and the QSLT on the coupling $\Omega$ and the cut-off frequency $\omega_{c}$ when $s=1$ and $\eta=0.9$. Fig.4(a) gives the non-Markovianity as a function of the coupling $\Omega $ for different values of $\omega_{c}$. If $\frac{\omega _{c}}{\omega _{0}}=2$, $\mathcal{N}$ is always zero when $\Omega<\Omega_{c}$ and $\mathcal{N}$ will increase with $\Omega$ enlarging when $\Omega>\Omega_{c}$. It should be noted that, if $\frac{\omega _{c}}{\omega _{0}}=1$ or $\frac{\omega _{c}}{\omega _{0}}=0.5$, the non-Markovian dynamics occurring for $\Omega=0.1\omega_{0}$ turns into Markovian and then back to non-Markovian by increasing $\Omega$, which such a behaviour has been also observed in different structured systems \cite{Man1,Man2}. But the critical value $\Omega_{c}$ is same for different values of $\omega_{c}$. Besides, we also find that, the smaller the value of $\frac{\omega _{c}}{\omega _{0}}$, the bigger the initial value of $\mathcal{N}$, and the bigger the value of $\mathcal{N}$ in areas with $\Omega>\Omega_{c}$. Fig.4(b) shows the QSLT as a function of the coupling $\Omega $ for different values of $\omega_{c}$. When $\frac{\omega _{c}}{\omega _{0}}=2$, $\tau_{QSLT}$ is always equal $\tau$ when $\Omega<\Omega_{c}$ and $\tau_{QSLT}$ will decrease with $\Omega$ enlarging when $\Omega>\Omega_{c}$. In particularly, when $\frac{\omega _{c}}{\omega _{0}}=1$ or $\frac{\omega _{c}}{\omega _{0}}=0.5$, $\tau_{QSLT}$ will increase from a certain value to one and then again quickly decrease from one with $\Omega$ enlarging and there is a same critical value $\Omega_{c}$ for different values of $\omega_{c}$. In addition, we can see that, the smaller the value of $\frac{\omega _{c}}{\omega _{0}}$, the smaller the initial value of $\tau_{QSLT}$, and the smaller the value of $\tau_{QSLT}$ in areas with $\Omega>\Omega_{c}$. Therefore, not only the atom-cavity coupling $\Omega$ but also cut-off frequency $\omega_{c}$ can enhance the non-Markovianity in the dynamics process and speed up the evolution of the atom.
	
	\begin{figure}[tbp]
		\includegraphics[width=7cm,height=4.5cm]{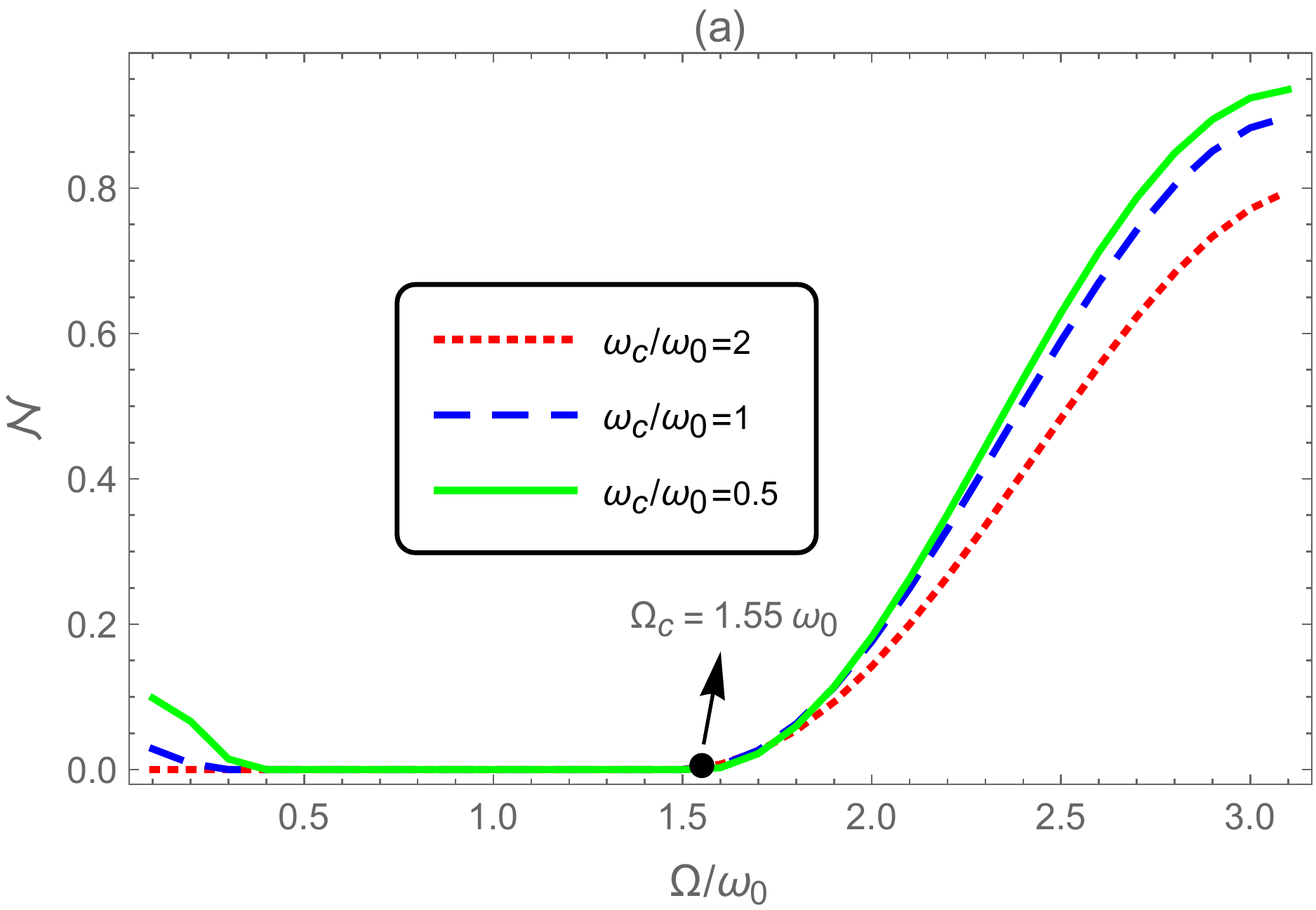}
		\includegraphics[width=7cm,height=4.5cm]{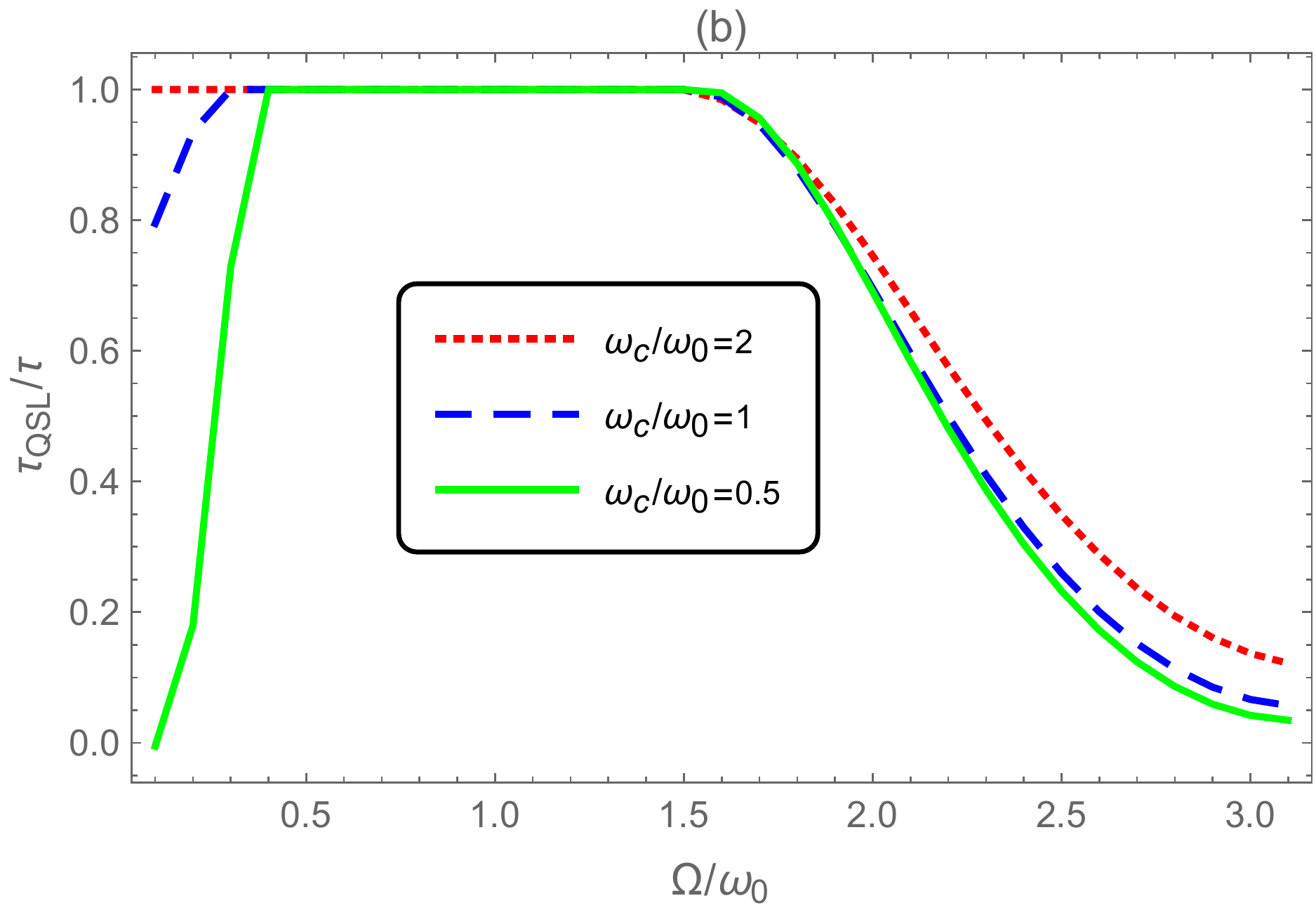}
		\caption{(Color online)Non-Markovianity and QSLT as a function of the coupling strength $\Omega$ when $s=1$ (Ohmic spectrum) for different cutoff frequency $\omega_{c}$, respectively. $\frac{\omega_{c}}{\omega_{0}}=2$, red dotted line; $\frac{\omega_{c}}{\omega_{0}}=1$, blue dashing line; $\frac{\omega_{c}}{\omega_{0}}=0.5$, red solid line. (a)Non-Markovianity as a function of $\Omega$; (b)QSLT as a function of $\Omega$. The other parameters are $\eta=0.9 $ and $\omega_{0}=1$.}
		\label{fig:4}
	\end{figure}
	
	The influences of $\Omega $ and $s$ on the non-Markovianity and the QSLT are shown in Fig.5. when $\eta=0.9$ and $\frac{\omega_{c}}{\omega_{0}}=2$. From Fig.5(a), we know that, for the Ohmic spectrum ($s=1$) and the super-Ohmic spectrum ($s=3$), $\mathcal{N}$ is always zero when $\Omega<\Omega_{c}$ and $\mathcal{N}$ will increase with $\Omega$ enlarging when $\Omega>\Omega_{c}$, and their critical values are different. However, for the sub-Ohmic spectrum ($s=\frac{1}{2}$), the non-Markovian dynamics occurring for $\Omega=0.1\omega_{0}$ also turns into Markovian and then back to non-Markovian by increasing $\Omega$, and the critical value under the sub-Ohmic spectrum is obvious less than that under the Ohmic spectrum. From Fig.5(b), we discover that, for the Ohmic spectrum ($s=1$) and the super-Ohmic spectrum ($s=3$), $\tau_{QSLT}$ is always zero when $\Omega<\Omega_{c}$ and $\tau_{QSLT}$ will decrease with $\Omega$ enlarging when $\Omega>\Omega_{c}$, and their critical values are different. However, for the sub-Ohmic spectrum ($s=\frac{1}{2}$), $\tau_{QSLT}$ will increase from 0.3 to one and then again quickly decrease from one with $\Omega$ enlarging, and the critical value under the sub-Ohmic spectrum is obvious less than that under the Ohmic spectrum. Namely, the atom-cavity coupling $\Omega$ and the Ohmicity parameter $s$ can effectively control the non-Markovianity in the dynamics process and speed up the evolution of the atom.
	
	\begin{figure}[tbp]
		\includegraphics[width=7cm,height=4.5cm]{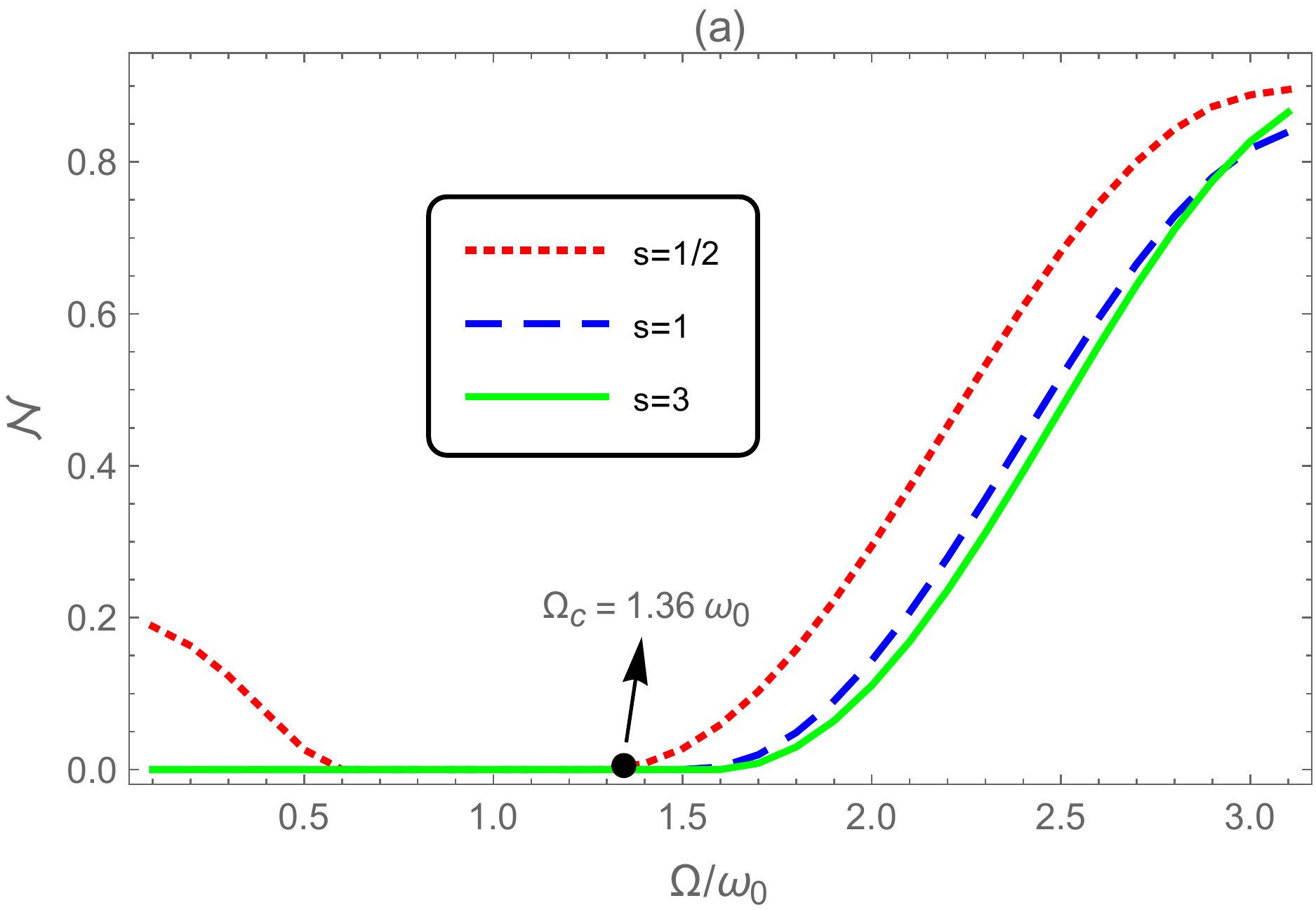} 
		\includegraphics[width=7cm,height=4.5cm]{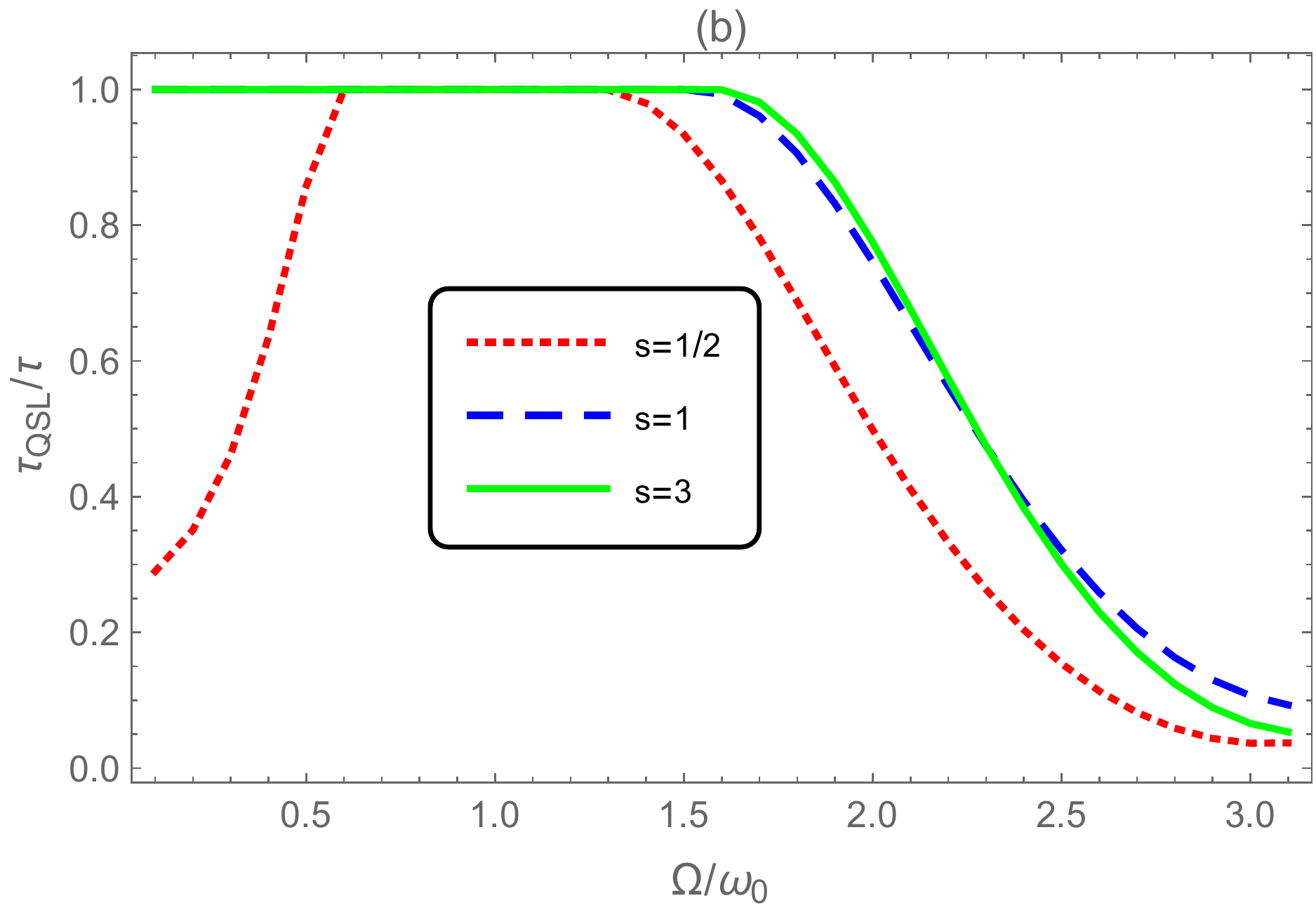} 
		\caption{(Color online)Non-Markovianity and QSLT as a function of the coupling strength $\Omega$ for different Ohmicity parameter $s$, respectively. $s=\frac{1}{2}$ (sub-Ohmic spectrum), red dotted line; $s=1$ (Ohmic spectrum), blue dashing line; $s=3$ (super-Ohmic spectrum), green solid line. (a)Non-Markovianity as a function of $\Omega$; (b)QSLT as a function of $\Omega$. The other parameters are $\eta=0.6$, $\omega_{0}=1$ and $\omega_{c}=2$.}
		\label{fig:5}
	\end{figure}
	
	In the following, the physical interpretation of the results above is given. Because the cavity coupling with the reservoir can be regarded as the environment of the atom, the energy and information can flow back from the environment to the atom through regulating the coupling strength $\Omega$. The larger the atom-cavity coupling $\Omega$, the more information the cavity flows back to the atom. Thus, the non-Markovianity will increase and the QSLT will decrease with $\Omega$ enlarging when $\Omega$ is bigger than the critical value. On the other hand, the influence of the cavity on the atom is obviously greater than that of the reservoir on the atom, so the critical value of sudden transition is mainly determined by $\Omega$. From Eq.~(\ref{EB208}), we know that a smaller value of $\eta$ corresponds to a longer correlation time of the reservoir thus the non-Markovianity $\mathcal{N}$ is bigger and the QSLT is smaller. Moreover, the smaller value of $\frac{\omega _{c}}{\omega _{0}}$ corresponds to the less overlap of the spectrum of the reservoir with the frequency of the cavity, that is, the reservoir is more effectively adiabatic and the non-Markovian effect is more obvious and the evolution of the atom is quicker. The smaller the Ohmicity parameter $s$ is, the smaller the peak and the width of the Ohmic spectral density are, the more obvious the non-Markovian effect is. So the smaller value of $s$ will lead to the larger non-Markovianity and the smaller QSLT. Besides, Eq.~(\ref{EB304}) shows that the information flows irreversibly from the atom to the environment so that the atom evolves at the actual speed and the QSLT is equal to the actual evolution time when $\mathcal{N}=0$. The information flows back from the environment to the atom thus the atom evolution is accelerated and the QSLT is smaller than the actual evolution time when $\mathcal{N}>0$.
	
	\section{Conclusion}
	In summary, we have investigated the non-Markovianity and the QSLT of the atom in Jaynes-Cummings model coupling with the Ohmic reservoir when the total excitation number is $n=1$. We have obtained the non-Markovianity characterized by using the probability of the atomic excited state and the negative decoherence rate in the time-local master equation (see Eq.~(\ref{EB3003})), which also showed that the non-Markovianity can be explained reasonably by the negative decoherence rate, namely, the dynamical process is non-Markovian if the decoherence rate is negative \cite{Hall2014a}. We have also studied in detail the influence of the atom-cavity coupling and the reservoir parameters on the non-Markovianity and the QSLT. The results have showed that, the atom-cavity coupling is the main physical reasons of the transition from Markovian to non-Markovian dynamics and the transition from no speedup to speedup process, and the critical value of this sudden transition only depends on the Ohmicity parameter. The appropriate reservoir parameters, such as the cavity-reservoir coupling $\eta$, the cut-off frequency $\omega_{c}$ and the Ohmicity parameter $s$, can improve the non-Markovianity in the dynamics process and speed up the evolution of the atom. In addition, we have also found that the non-Markovian dynamics occurring for $\Omega=0.1\omega_{0}$ turns into Markovian and then back to non-Markovian by increasing $\Omega$ when $\frac{\omega _{c}}{\omega _{0}}=1$, $\frac{\omega _{c}}{\omega _{0}}=0.5$ or $s=\frac{1}{2}$( the sub-Ohmic spectrum). 
	
	In this work, only zero temperature reservoir is considered. If the reservoir is at nonzero temperature, from Eq.~(\ref{EB2031}), we can see that, the quantum coherence of the atom-cavity and the populations of the states $|\varphi _{1,\mp}\rangle$ will increase a little under the effect of thermal reservoir. The non-Markovianity and the QSLT of the atom will be different from zero temperature case. The detailed influence of nonzero temperature on quantum effect will be presented in our next work. These results will provide interesting perspectives for future applications of open quantum systems in quantum physics \cite{Varcoe,Jonathan,Nori,Prawer}.
	
	\begin{acknowledgments}
		This work was supported by the National Natural Science Foundation of China (Grant No 11374096) and the Doctoral Science Foundation of Hunan Normal University, China.
	\end{acknowledgments}

\end{document}